# Simulation of wireless dynamic source routing protocol with IP traffic flow analysis, memory efficiency and increased throughput

Diya Naresh Vadhwani[1], Megha Singh[2], and Deepak Kulhare[3]
[1] Central India Institute of technology, Indore, India
dnvadhwani@gmail.com
[2] Central India Institute of technology, Indore, India
maggii.megha@gmail.com,
[3] Central India Institute of technology, Indore, India
dkulhare@gmail.com

*Abstract*—Today in fast technology development in wireless mobile adhoc network there is vast scope for research. As it is known that wireless communication for mobile network has many application areas like routing services, security services etc. The mobile adhoc network is the wireless network for communication in which the mobile nodes are organized without any centralized administrator. There are many Manet routing protocols like reactive, proactive, hybrid etc. In this paper the reactive Manet routing protocol like DSR is simulated for traffic analysis for 50 mobile nodes for IP traffic flows. Also throughput is analyzed for DSR and ER-DSR protocol. And finally the memory utilized during simulation of DSR and ER-DSR is evaluated in order to compare both.

## I. INTRODUCTION

The wireless network has many advantages in today technological development for communication of wireless and mobile nodes. The wireless network is of types types, first the wireless network with base station or access point and second the wireless network without any base station which is also know as ad hoc network. In this paper the mobile ad hoc network is studied for routing in order to simulate the dynamic source routing protocol for various performance matrices. Here during the simulation the IP traffic flows are also analyzed.

Also the ER-DSR which is the enhanced DSR is implemented and simulated in OPNET 14.5 simulator. Finally the results for throughput and memory utilization are shown to compare the original DSR and ER-DSR.

## II. DSR INTRODUCTION

DSR is the dynamic reactive source routing protocol which is based on source routing and does not based on based. DSR possessed the two mechanism for routing first is "Route discovery" and second is "Route maintenance".
DSR accumulates address of each device during route discovery process [1]. DSR uses source routing process and does not use hop by hop routing. Source routing is a routing technique in which the sender of a packet determines the complete sequence of nodes through which to forwarding "hop" by the address of the next node to which to transmit the packet on its way to the destination node [1].
Here in this research work the dsr manet routing protocol is updated which is known as ER-DSR using the dsr route discovery parameters and the simulation and results are collected.

## III. SIMULATION ENVIRONMENT

Here the dynamic source protocol and ER-DSR are simulated using OPNET 14.5 modeler and the results for memory used by protocol during simulation, number of packets delivered per unit time i.e. throughput and IP-traffic flow analysis are collected and shown with simulations.
Fig.1 shows the network diagram for 50 nodes.

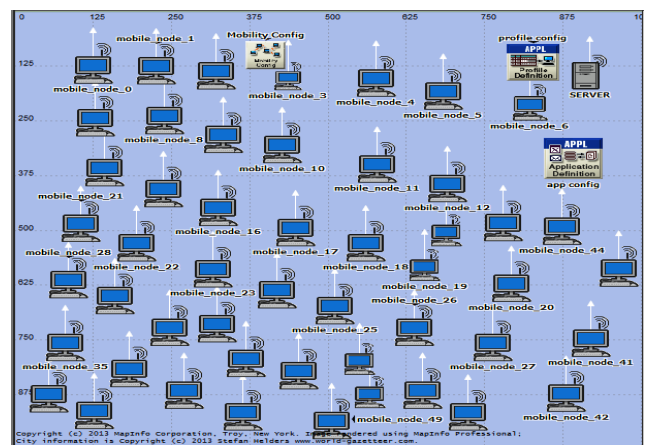

Fig 1.Network Diagram for 50 nodes

Fig 2 shows the node model for wireless mobile node.

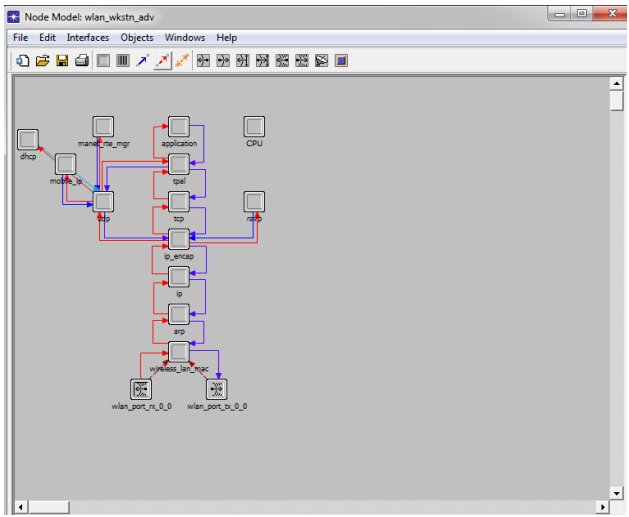

Fig 2.Node model for mobile node

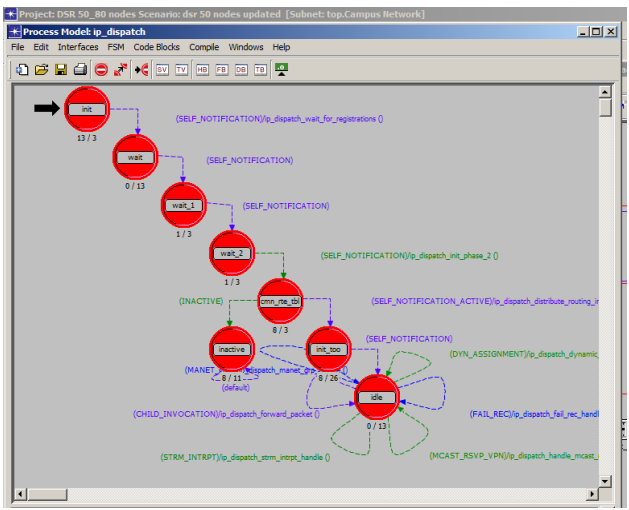

Fig 3 Process model for ip_dispatch

## IV. IP TRAFFIC ANALYSIS FOR PROPOSED SCHEME

*A. Traffic Analysis for proposed protocol*

The traffic analysis parameters for the DSR routing protocol are as shown in Table I.

TABLE I . TRAFFIC PARAMETERS

| Parameter | Value |
|---|---|
| Traffic Type | IP-unicast traffic |
| Number of Nodes | 50 mobile nodes |
| Channel | Wireless |
| Protocol | DSR & ER-DSR |
| Simulation time | 360 sec |
| Transmit power | 0.005W |
| Traffic Protocol | IP |
| Traffic Bit/sec | 12000 |

| Parameter | Value |
|---|---|
| Traffic Type | IP-unicast traffic |
| Traffic packets/sec | 100 |

The IP-traffic flows are shown in Table II as below.

TABLE II . IP-TRAFFIC FLOWS

| From Mobile Node 1 to Mobile Node 48 | When the simulation Starts |
|---|---|
| From Mobile Node 1 to Mobile Node 33 | Starts after 60 sec of simulation |
| From Mobile Node 1 to Mobile Node 20 | Starts after 60 sec of simulation |
| From Mobile Node 1 to Mobile Node 41 | Starts after 60 sec of simulation |
| From Mobile Node 1 to Mobile Node 46 | Starts after 60 sec of simulation |
| From Mobile Node 1 to Mobile Node 10 | Starts after 60 sec of simulation |
| From Mobile Node 1 to Mobile Node 44 | Starts after 60 sec of simulation |

The Socket information is shown in Table III.

TABLE III.SOCKET INFORMATION

| Destination IP address | Auto assigned |
|---|---|
| Type of service | Best Effort(0) |
| Protocol | IP |
| Source | Email |
| Destination | Ftp-Data |

*B. IP Traffic Flows*

The Fig 4 shows IP traffic flows for ER-DSR Manet routing protocol.

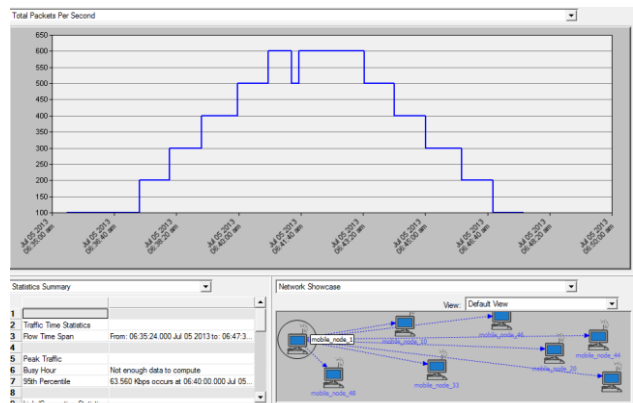

Fig 4 IP Traffic flow analysis for ER-DSR

The Fig 5 shows IP traffic flows for DSR Manet routing protocol

## V. THROUGHPUT COMPARISON

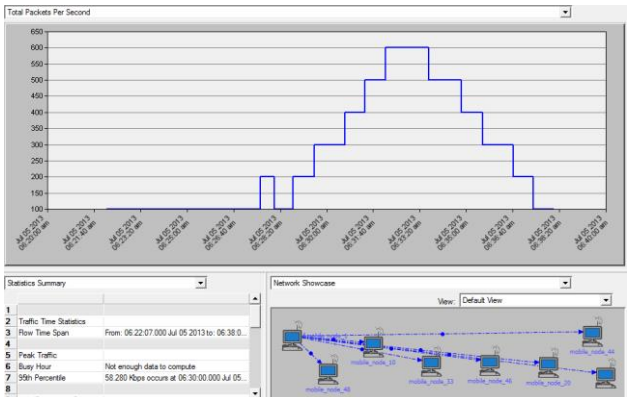

Fig 5 IP Traffic flow analysis for DSR

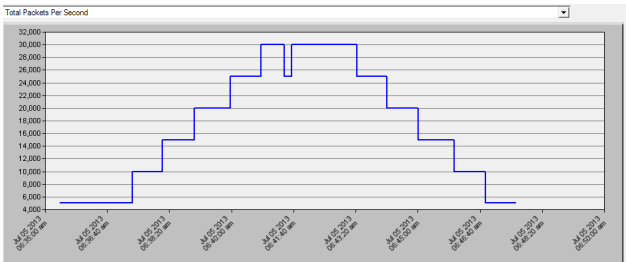

Fig 6 Total Traffic flow analysis for ER DSR

TABLE IV. Overall Traffic Summary statistics

| Parameter | Value |
|---|---|
| Number of Flows | 350 |
| Total volume on flows | 180.244MB |
| Average volume per flow | 527.344KB |
| Total volume | 180.244MB |
| IP Protocol | 180.244MB in 350 flows |
| Flow Average | 15.000B |

TABLE V. Traffic Summary statistics for 7 flows defined

| Parameter | Value |
|---|---|
| Number of Flows | 7 |
| Total volume on flows | 3.605MB |
| Average volume per flow | 527.344KB |
| Total volume | 3.605MB |
| IP Protocol | 3.605MB in 7 flows |
| Flow Average | 15.000B |

TABLE VI. Peak Traffic Summary statistics for DSR and ER-DSR

| Peak traffic | 95[th] Percentile value |
|---|---|
| For ER-DSR | 63.560 Kbps |
| For DSR | 58.280 Kbps |

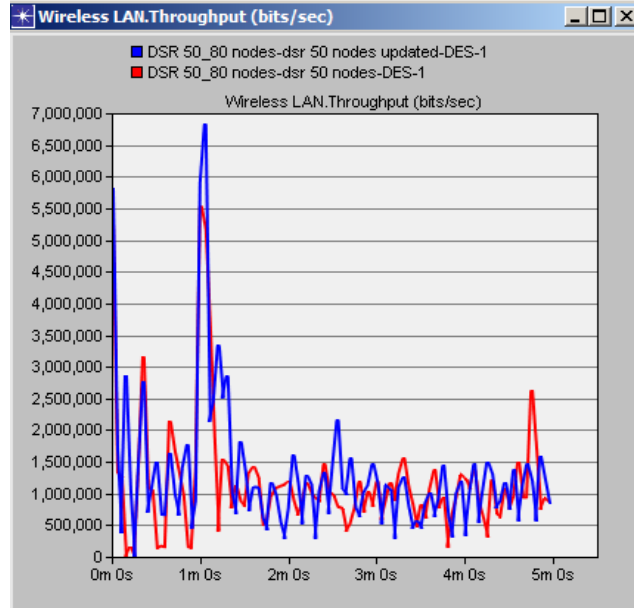

Fig 7 Total throughput comparison for DSR and ER-DSR

TABLE VII. Statistics summary for throughput for DSR and ER-DSR

| Parameters | ER-DSR Throughput (bits/sec) | DSR Throughput (bits/sec) |
|---|---|---|
| Initial value | 5,819,312 | 5,280,896 |
| Standard deviation | 1,055,142.93989592 | 928,386.914082489 |
| 99% conf-interval value | 993,836.108728629 | 898,084.458036654 |

## VI. MEMORY UTILIZATION

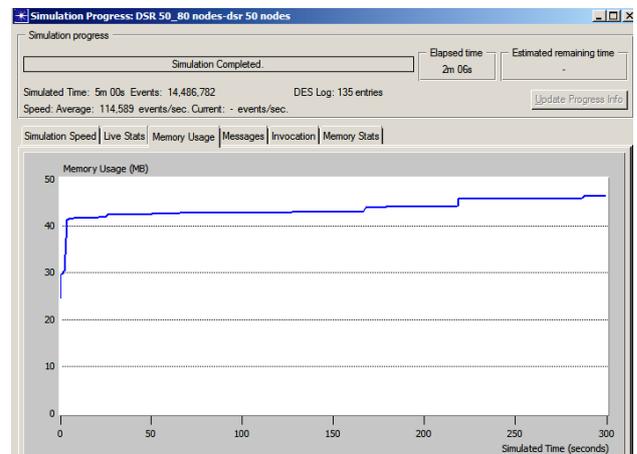

Fig 8 Total Memory used for DSR

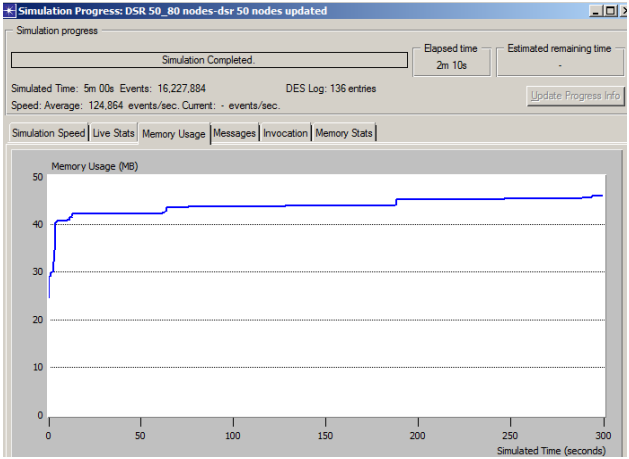

Fig 9 Total Memory used for ER-DSR

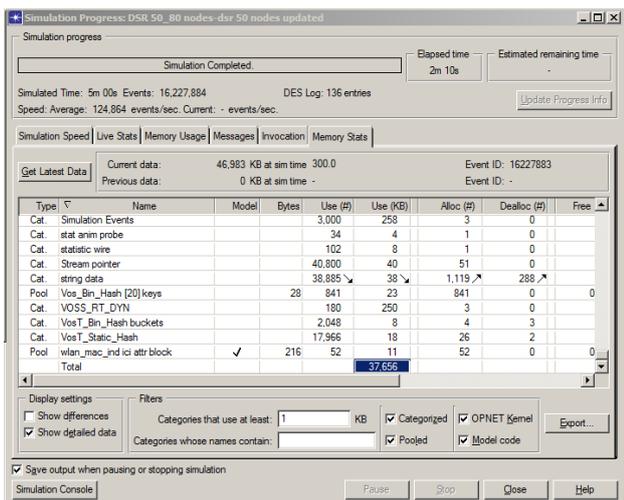

Fig 10 Memory statistic used for ER-DSR

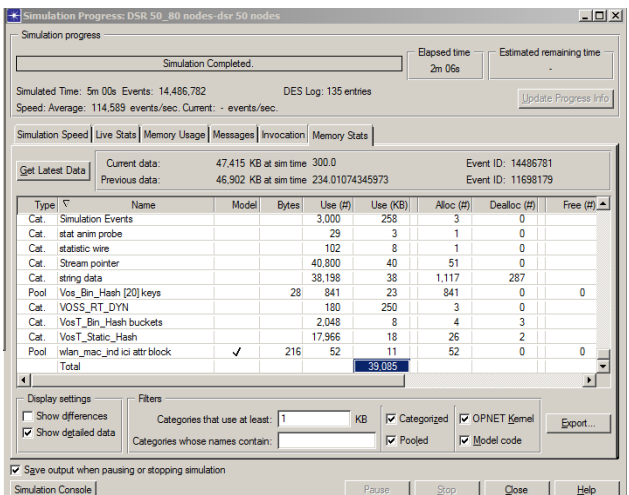

Fig 11 Memory statistic used for DSR

TABLE VIII. Statistics summary of Memory usage for DSR and ER-DSR

| Parameters | ER-DSR | DSR |
|---|---|---|
| Memory used (Mean value (KB)) | 37,656 KB | 39,085 KB |
| Speed (events/sec) | 124,864 events/sec | 114,589 events/sec |
| Simulated time (sec) | 300 sec | 300 sec |

CONCLUSIONS

From the above all figures 1 to 9 the DSR and ER-DSR routing protocols are analyzed for traffic flow analysis, throughput and memory utilization which are shown with graph.

From Table IV and Table V and VI it is concluded that total volume on 350 flow is 180.244 MB, average volume per flow is 527.344 KB and total volume on 7 flows is 3.605MB. Finally the peak traffic for ER-DSR is 63.560 Kbps and for DSR 58.280Kbps.

From Table VII the throughput variations are observed. For DSR the throughput is 898,084.458036654bits/sec while for updated or enhanced DSR (ER-DSR) the throughput value is increased as 993,836.108728629 bits/sec. Hence it is concluded that ER-DSR proposed scheme can increased the throughput value by 9.5751%, but not always it can be vary depending on node size.

From Table VIII, the memory used for DSR simulation is 39,085KB while the memory used for ER-DSR simulation is 37,656 KB .Thus the ER-DSR is made efficient in terms of memory usage. Also the simulation speed is 124,864 events/sec for ER-DSR and for DSR is 124,864 events/sec.

For future the DSR can be made more efficient and reliable by considering the another performance matrices.

**Authors Profile** :

1. Diya Naresh Vadhwani
   MTECH Student of CSE, CIIT, RKDF group of institution Indore, RGPV University Madhya Pradesh, India

2. Megha singh
   Asst. Prof of Department of CSE, CIIT, RKDF group of institution Indore, RGPV University Madhya Pradesh, India

3. Prof. Deepak Kulhare
   Head of the Department of CSE, CIIT, RKDF group of institution Indore, RGPV University Madhya Pradesh, India